\documentclass{article}

\usepackage{PRIMEarxiv}

\usepackage[utf8]{inputenc} % allow utf-8 input
\usepackage[T1]{fontenc}    % use 8-bit T1 fonts
\usepackage{hyperref}       % hyperlinks
\usepackage{url}            % simple URL typesetting
\usepackage{booktabs}       % professional-quality tables
\usepackage{amsfonts}       % blackboard math symbols
\usepackage{nicefrac}       % compact symbols for 1/2, etc.
\usepackage{microtype}      % microtypography
\usepackage{lipsum}
\usepackage{fancyhdr}       % header
\usepackage{graphicx}       % graphics
\usepackage{bm}
\usepackage{amsmath}
\usepackage{bm}
\usepackage{array}
\usepackage{subfigure}
\usepackage{xcolor}
\graphicspath{{media/}}     % organize your images and other figures under media/ folder

\DeclareMathAlphabet\mathbfcal{OMS}{cmsy}{b}{n}
\newcommand{\rr}{{\bm{r}}}
\newcommand{\md}{{\rm{d}}}

\newcommand{\mi}{{\rm{i}}}
\newcommand{\ms}{ms$^{-1}$}

\title{Atomic diffraction by patterned holes in hexagonal boron nitride: a comparison between semi-classical and quantum computational models.
}

\author{
  E. K. Osestad,\\
  Department of Physics and Technology \\
  University of Bergen \\
  Bergen\\
  \texttt{eivind.osestad@student.uib.no} \\
   \And
  E. Zossimova \\
  Freiburg Center for Interactive Materials and Bioinspired Technologies (FIT) \\
  University of Freiburg\\
  Freiburg\\
   \And
  M. Walter \\
  Freiburg Center for Interactive Materials and Bioinspired Technologies (FIT) \\
  Cluster of Excellence livMatS @ FIT \\
  Fraunhofer IWM, MikroTribologie Centrum {$\mu$}TC\\
  University of Freiburg\\
  Freiburg\\
   \And
  J. Fiedler \\
  Department of Physics and Technology \\
  University of Bergen \\
  Bergen\\
}

\begin{document}
\maketitle

\begin{abstract}
The diffraction of atoms and molecules through tiny, sub-nanometre holes in atomically thin membranes is a promising approach for advancing atom interferometry sensing and atomic holography. However, dispersion interactions, such as the Casimir-Polder force, pose a significant challenge by attracting diffracting particles to the membrane, limiting the minimum hole size. This paper presents a numerical simulation of helium matter-wave diffraction through sub-nanometre holes in hexagonal boron nitride by solving the time-dependent Schrödinger equation. Our results show that the transmission rates in the quantum approach are significantly higher than those predicted by the commonly used semi-classical approach. This suggests that significantly smaller holes can be used in the design of diffractive masks, provided that fabrication techniques can meet the atomic-level precision to realise such holes. Furthermore, we observe notable differences in diffraction patterns, even for atom velocities that are much greater than the expected convergence threshold between semi-classical and quantum computational models.
\end{abstract}

Diffractive masks with extremely small, sub-nanometre-sized holes offer a way to control atomic wavefronts with high precision. This control enables the creation of high-sensitivity interferometers for more precise measurements of fundamental physical constants{, as smaller holes impart a larger momentum transfer,}~\cite{Tino_2021, Keith1991, Parker2018} and inertial sensing~\cite{PhysRevLett.116.183003,PhysRevA.108.023306,Saywell2023}. Additionally, the ability to manipulate atomic waves with high precision could enable high-resolution atom holography~\cite{Nesse17, Nesse19, fujita1996manipulation,Fiedler_2023}.

However, the minimum size of the hole that can be used for atomic diffraction depends on the strength of the dispersion forces, such as Casimir, Casimir--Polder and van der Waals forces~\cite{WIPS,doi:10.1142/4505,doi:10.1142/9383, Cronin2005_1,PhysRevB.101.235424,https://doi.org/10.1002/andp.201500224, Osestad2025}, arising from the ground-state fluctuations of the electromagnetic field between incident atoms and edge atoms around the hole in the diffraction mask. Attractive dispersion forces can curve the trajectory of incident atoms, reducing the transmission area of the hole~\cite{Buhmann12a, Brand15, Fiedler2022, PhysRevLett.83.1755}. These forces also shift the phase of atomic waves passing through the hole, which affects the resulting diffraction patterns~\cite{Fiedler2022, Cronin2005_2}. The strength of dispersion forces depends on multiple factors, including the thickness of the diffraction mask; the atomic number and polarisability of the edge atoms in the mask; the distance between the incident atoms and the edge atoms; as well as the velocity and properties of the incident atoms.

Dispersion forces scale with the thickness of the material used for the diffraction mask (i.e. the depth of the hole). Therefore, atomically thin materials can be used to minimise dispersion forces and design diffraction masks with sub-nanometre-sized holes. Monolayer materials, such as graphene or hexagonal boron nitride (hBN), are suitable candidates, due to their chemical stability and mechanical properties~\cite{Roy2021}. The incident atoms that are shot into the diffraction mask should be small atoms, such as hydrogen and helium, since their comparatively low number of electrons {and internal states} makes their dispersion interactions with the atoms in the diffraction mask weaker.

In a previous study, we have theorised that it should be possible to diffract helium atoms by sub-nanometre holes in {hBN}, when the atoms are propagating at normal incidence through the hole at high velocities (on the order of $20$\,{k}\ms,  equivalent to 8.3\,eV kinetic energy)~\cite{Osestad24}. Furthermore, others have shown that it {is possible to} diffract {molecules} through holes in graphene~\cite{Brand2015}. Recent experiments have further demonstrated that natural pores in the graphene lattice can act as a diffraction grating for hydrogen and helium atomic waves, when the atoms transmit at normal incidence to the lattice plane with velocities {greater} than $150$\,{k}\ms~{(or about 466\,eV of kinetic energy)}\cite{kanitz2024, Brand_2019}. {For lower velocities, the repulsive forces between atoms at close range lead to scattering in the other direction\cite{Beeby_1971, celli1985, Farias_1998, PhysRevLett.75.886}.}

From a computational perspective, there are different methods to calculate the propagation of atomic waves through {patterned holes or slits in materials}, including the resulting diffraction patterns and phase shifts\cite{Brand_2019, Brand2015, Fiedler2022,PhysRevLett.125.050401}. 
The incident matter waves can be described using either {semi-}classical or quantum mechanics, whilst the electronic properties of the interacting atoms are described using electronic structure theory. {A semi-classical approximation holds when the change in de Broglie wavelength over distance is much less than 1\cite{messiah1961quantum}. As there are sharp spikes in the potentials resulting from dispersion interaction, it is difficult to say exactly when they hold.} 
%A semi-classical approximation is typically valid when the average diameter of the hole is larger than the de-Broglie wavelength of the incident matter wave\. For holes that have a diameter of $\sim 1$\,nm, the semi-classical model should provide a good approximation at velocities above approximately $40$\,{k}\ms\, for hydrogen and $17$\,{k}\ms\, for helium{, which correspond to 8.3 and 6.0eV of kinetic energy respectively}. These can be considered as the thresholds at which the classical and quantum models start to converge. 
Close to these thresholds, the two descriptions can yield different results\cite{Garcion2024}. 

In this paper, we first use electronic structure theory to calculate the dispersion forces between incident helium atoms and atoms in the hBN monolayer. Then, we introduce a numerical method to solve the propagation of the transverse part of the wavefunction as it passes through different holes in hBN. We compare the results with our previous semi-classical calculations~\cite{Osestad24} and find that there are notable differences, especially at lower velocities, where the de-Broglie wavelength of the matter wave is comparable to the size of the hole. {We limit ourselves to modelling coherent diffraction. Effects, such as phonon excitation in the monolayer, are ignored\cite{Jiang2019, Brand_2019}.} One key difference is that, within the quantum description, the atoms can pass through smaller holes at lower velocities than previously thought possible.

% In this paper, we will first describe electronic structure theory to find the electromagnetic properties of the atoms in the monolayer, which is needed to describe the electromagnetic and dispersion forces acting between the diffracting atom and the monolayer holes. Then, we will introduce a numerical method and solve the propagation of the transverse part of the wavefunction as it passes through the hole. 

\section{Computational approach} 
\begin{figure}[t]
    \centering
    \includegraphics[width=0.5\linewidth]{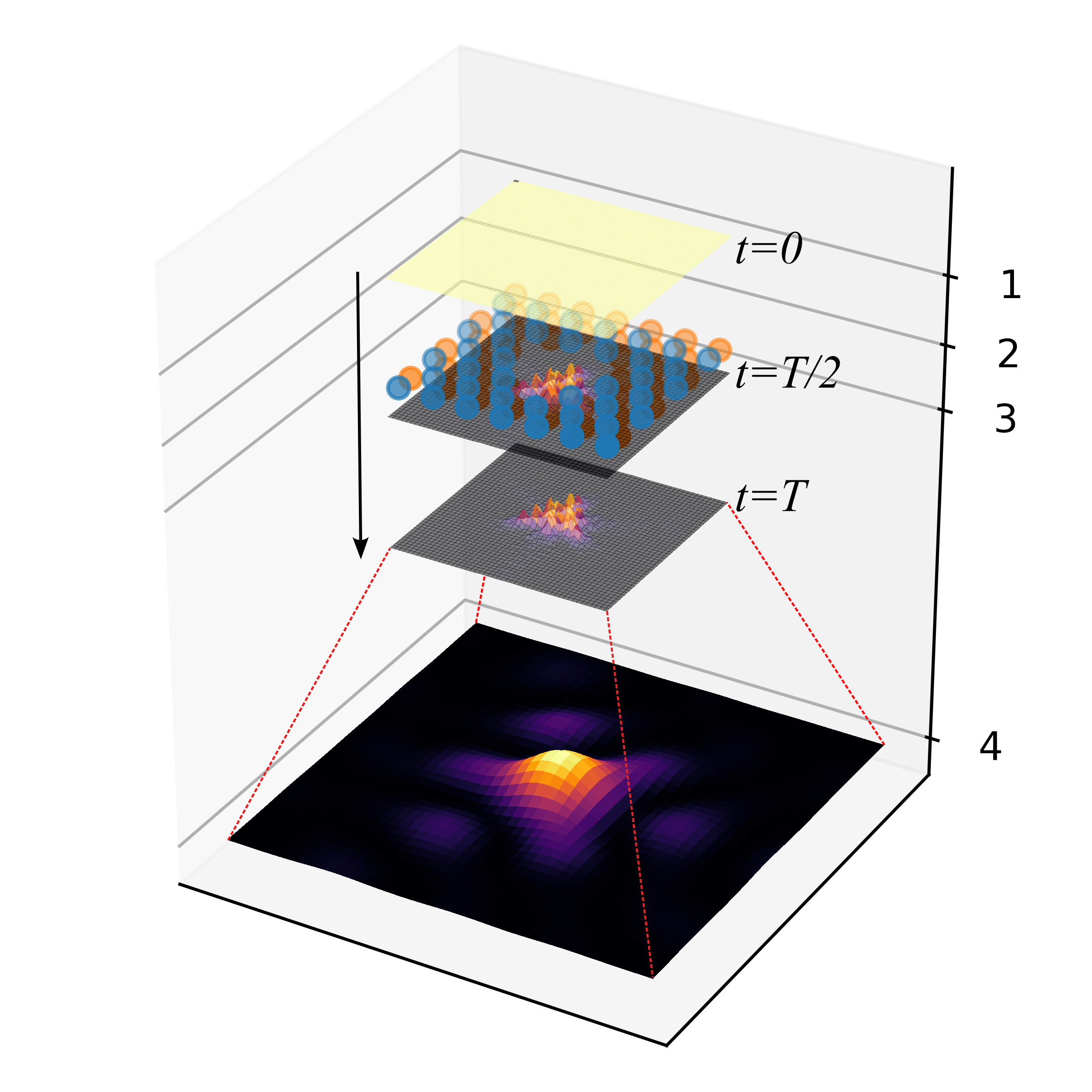}
    \caption{Schematic of the simulation setup, with some of the intermediate steps shown. {We use the paraxial approximation, where the $z$-coordinate evolves as $z(t) = v t + z_0$, while the $x$ and $y$ coordinate is used in the numerical Scrödinger equation simulation. We simulate a $15.9\times15.9$Å grid in the $xy$-plane. }{Layers 1-3 are snapshots of the simulation at different points during the full simulation time $T$.} Layer 1 shows the initial simulated state of the wavefunction. It has a uniform distribution over the simulated plane. The simulated plane propagates downwards through the hole in the hBN monolayer, where the atoms are represented by blue and orange spheres {around} Layer 2. Layer 3 shows the final simulated plane {at time $t=T$. The diffraction pattern is calculated from this state}. The resulting diffraction pattern in the observation plane (Layer 4) is calculated using the Kirchhoff diffraction formula in Eq.~\eqref{eq:kirchhoff}. The example diffraction pattern in Layer 4 is not to scale.}
    \label{fig:setup-illustration}
\end{figure}

Our computational approach involves simulating the propagation of a helium matter wave through a hole in a hBN monolayer. The simulation setup, illustrated in Fig.~\ref{fig:setup-illustration}, consists of four conceptual layers. Layer 1 represents the initial homogeneous wave function on a plane parallel to the hBN monolayer. As the matter wave propagates downward through the hBN plane in Layer 2, we capture snapshots of the wave function at each time step.

\begin{figure*}
    \centering
    \subfigure[6\,Å hole]{\includegraphics[width=0.27\textwidth]{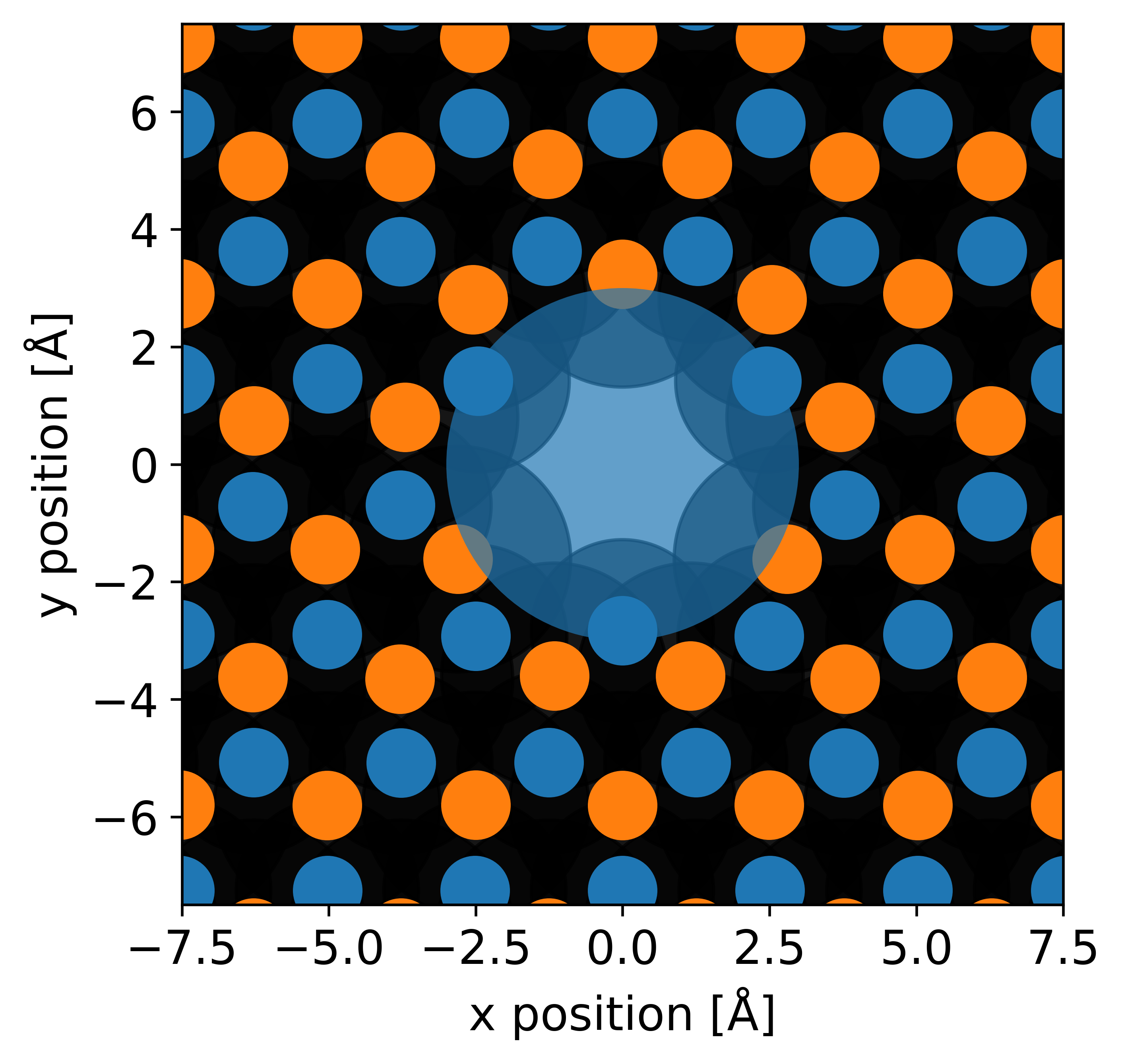}}
    \subfigure[10\,Å hole]{\includegraphics[width=0.27\textwidth]{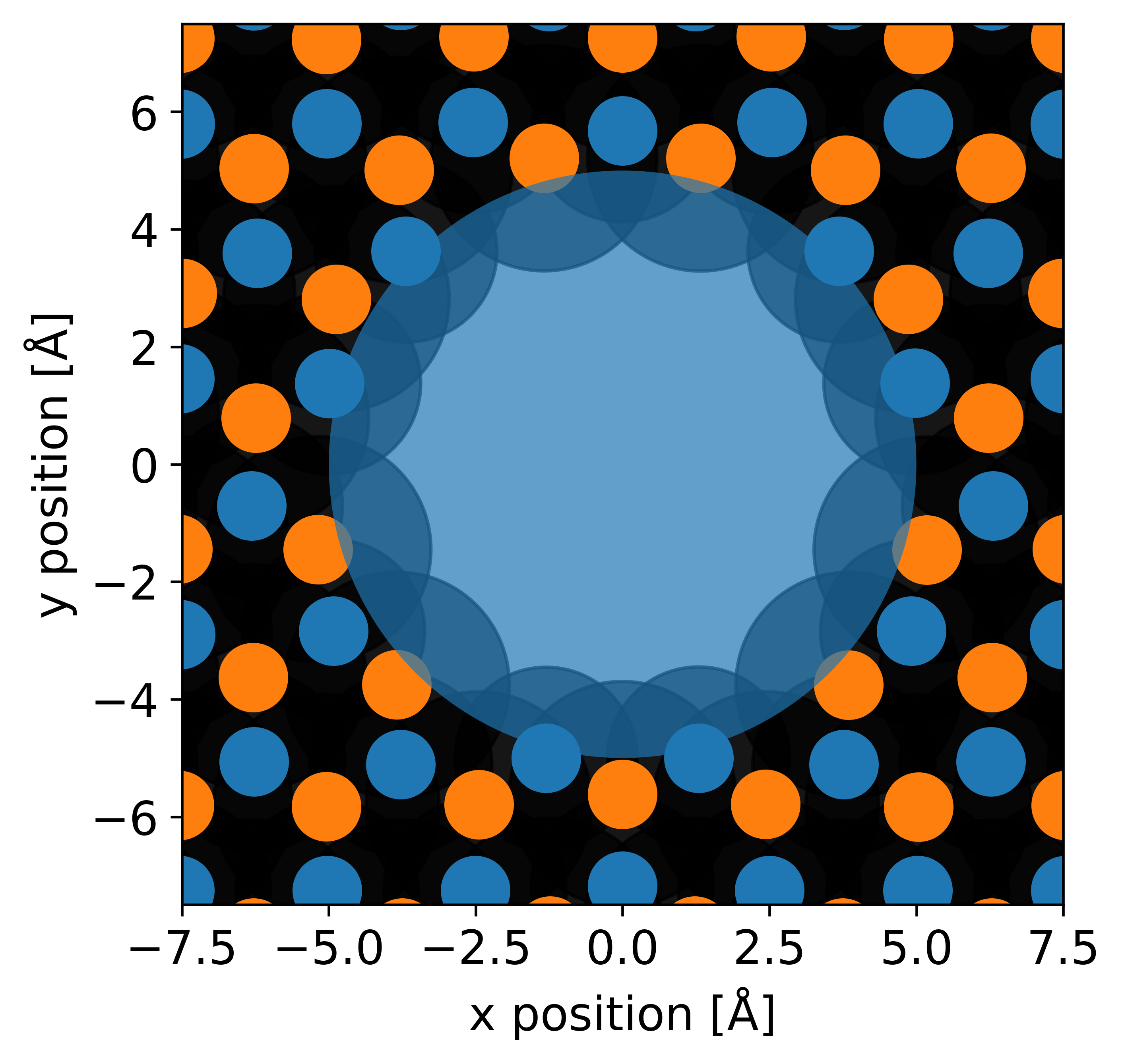}}
    \subfigure[snowflake hole]{\includegraphics[width=0.27\textwidth]{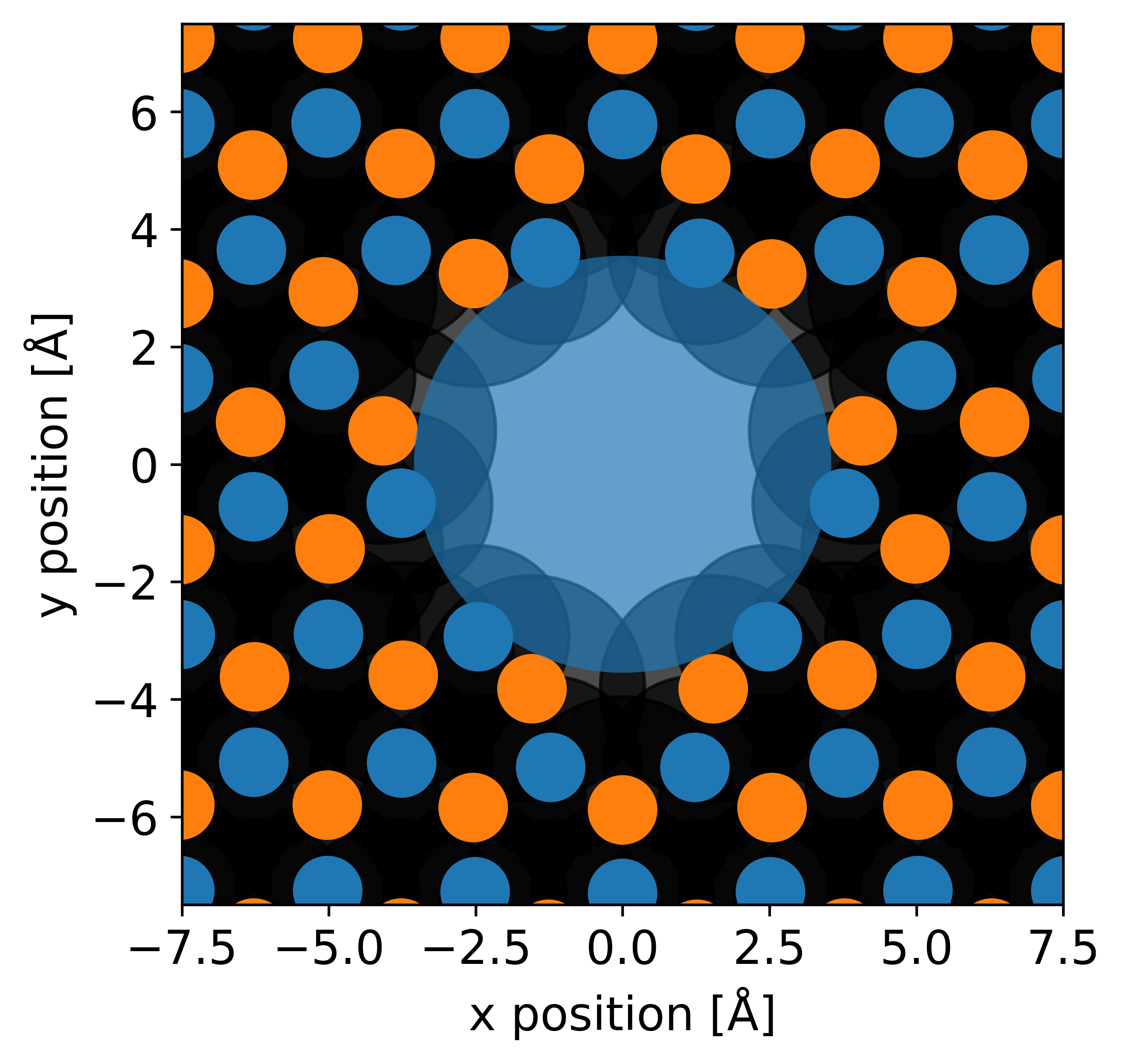}}
    \caption{The shape of the three holes we consider, with nitrogen atoms in blue, and boron atoms in orange. {The {central blue shaded circles} are used to compare the transmission at different velocities. {The dark circles around the atoms shows the absorption area of the atoms.} The {central blue shaded circles} are 28.3\,Å$^2$ for the 6\,Å hole, (b) 78.5\,Å$^2$ for the 10\,Å hole and 39.6\,Å$^2$ for the snowflake hole. {The central blue circles are meant only for comparison, and does not represent the apparent size in any model.}}
    }
    \label{fig:holes_shape}
\end{figure*}

The hBN monolayer is modelled using a 12x12 supercell of the hBN primitive cell, with a lattice constant $a_{\rm l}=$ {{$ 2.504\,\rm{Å}$}}\cite{C7RA00260B,doi:10.1080/10584587.2015.1039410}. We create hBN supercells with 3 different hole geometries, as depicted in Fig.~\ref{fig:holes_shape}, with the help of the atomic simulation environment\cite{larsen_atomic_2017}. These include circular holes with diameters of 6\,\AA\ (a) and 10\,\AA\ (b), as well as a snowflake-shaped hole (c). Each of these hBN supercells is considered individually in Layer 2, allowing us to study the impact of hole geometry on the matter wave propagation.

As the matter wave interacts with the hBN monolayer, a significant portion of the wave front is scattered, with most particles being back-scattered, while allowing some to transmit through the monolayer. The transmitted part of the wavefront in Layer 3 evolves freely, eventually forming an interference pattern, as shown in Layer 4. Note that the labelling of layers is a conceptual framework for understanding the simulation, rather than the time-step resolution of the numerical simulation.

%Fig.~\ref{fig:setup-illustration}. Finally, we will demonstrate expected diffraction patterns, %This simulation of the wavefront can then be used to find the far field diffraction pattern. We then 
%which we will compare to the patterns obtained via the semi-classical eikonal approximation.

% The setup considered is illustrated in Fig.~\ref{fig:setup-illustration}. The matter wave propagates downwards, which will be treated as multi-slice propagation, meaning the time-evolution occurs in transverse planes labelled as layers. Layer 1 denotes the homogeneous initial matter wave. The membrane is located in layer 2, where most of the wave front is absorbed. Here, absorption considers the incoherently scattered amount of particles. Layer 3 indicates the free evolution of the transmitted wave which creates the interference pattern in layer 4. Note that the labelling of layers is not discretisation for the actual simulation.

%We find that it is possible to pass the holes at much slower speeds than calculated with our previous method\cite{Osestad24}, and that the eikonal approximation is not sufficient at these lower speeds. 

\subsection{Electronic structure theory}
The dispersion interactions between the hBN monolayer and incident helium atoms depend on the electronic properties of the interacting atoms, namely their polarisabilities. Atomic polarisability tensors can be obtained from electronic structure theory via density functional theory (DFT)\cite{mortensen_gpaw_2024}. Here, we build upon our previous results~\cite{Osestad24} with the computational settings detailed there.
In brief, DFT is applied, where the exchange–correlation energy
is described by the PBE functional\cite{Perdew96prl}. Tkatchenko–Scheffler dispersion corrections\cite{tkatchenko_accurate_2009, tkatchenko_accurate_2012}
are used to determine the screened atomic polarisabilities as input for our quantum description of helium atom diffraction by holes in hBN.

%{\color{blue}JF: instead of rewriting everything, we can describe what we did and refer to our previous paper}
%\mw{I agree, no equations needed from our side.}

\subsection{Dispersion Interactions}
The main long-range interaction between the monolayer and the diffracted atoms is the Casimir--Polder interaction\cite{Buhmann12a,doi:10.1142/9383,Scheel2008}
\begin{equation}
    U_{\rm CP}(\rr) = \frac{\hbar\mu_0}{2\pi}\int\limits_0^\infty \md \xi\, \xi^2\operatorname{Tr}\left[\boldsymbol{\alpha}(\mi\xi)\cdot {\bf{G}}(\rr,\rr,\mi\xi)\right]\,,\label{eq:CP}
\end{equation}
where $\hbar$ is the reduced Planck constant; $\mu_0$ is the vacuum permeability; $\bf{G}$ is the Green's function containing the properties of the monolayers; and $\boldsymbol{\alpha}$ the frequency-dependent polarisability describing the properties of the diffracted atom. $\xi$ is an imaginary frequency number. The interaction can be viewed as the sum of all virtual photons originating at the diffracting atom, travelling to the monolayer and then returning to the atom. We can approximate the scattering process of virtual photons by the first-order Born series. {Since hBN is a insulator with a weak dielectric response, higher orders should be negligible~\cite{Buhmann12b}}. In this approach, the Green function can be written as the combination of propagations to and from each atom in the monolayer\cite{WIPS,C9CP03165K} 
\begin{equation}
{\bf{G}}({\bm{r}},{\bm{r}}',\omega) = \frac{\omega^2}{c^2\varepsilon_0}\sum_i{\bf{G}}({\bm{r}},{\bm{r}}_{i},\omega)\cdot \boldsymbol{\alpha}_i(\omega) \cdot {\bf{G}}({\bm{r}}_i,{\bm{r}}',\omega)\,, \label{eq:Born}
\end{equation}
where $\boldsymbol{\alpha}_i$ and $\bm{r}_i$ are the screened polarisability 
(determined by DFT) and position of the $i$-th monolayer atom. This means that, in our case, the Casimir--Polder force between the monolayer and the diffracting atom is the sum of van der Waals potentials between the atoms in the monolayer and the diffracting helium atom.

%, given by
%\begin{equation}
%    U_{\rm vdW}(\rr) = - \frac{C_{6,\, j}}{\left|\rr\right|^6} \, .
%\label{eq:vdw}
%\end{equation}

Inserting Eq.~\eqref{eq:Born} into Eq.~\eqref{eq:CP}, leads to\cite{Osestad24}
\begin{equation}
\begin{split}
    U_{\rm CP}(\bm{r}) = & -\sum_i \frac{C_6^{(i)}}{6|\bm{r}_i - \bm{r}|^6} \\ 
      \cdot & \left[{\rm Tr} \boldsymbol{D}_i + 3\frac{ (\bm{r}_i - \bm{r}) \cdot\boldsymbol{D}_i\cdot (\bm{r}_i - \bm{r})}{|\bm{r}_i - \bm{r}|^2}\right] \,,
\end{split}
\label{eq:Uvdw}
\end{equation}
where $\boldsymbol{D}_i$ is a $3\times3$-matrix encoding the directional dependence of the polarisability, $\boldsymbol{\alpha}_i(\omega) = \alpha_i(\omega) \cdot \boldsymbol{D}_i$. {The values of $C_6^{(i)}$ and $\boldsymbol{D}_i$ is taken from Ref.~\cite{Osestad24}. The $|\bm{r}_i -\bm{r}|^{-6}$ dependence ensures convergence of the sum from Eq.~\eqref{eq:Born}. This is essentially a discrete version of the Hamaker approach~\cite{HAMAKER1937}. If we had a solid material with a well-defined thickness, we would get an integral resulting in a $1/r^3$ dependent potential. This is not feasible with monolayers, as their thickness is difficult to define.} 

\subsection{Electrostatic potential}
As an insulator, hBN exhibits a distribution of partial charges, where each atom is assigned an effective charge. While these charges are neutralised on large scales, they give rise to a local electric field near the monolayer or hole. This field, in turn, induces a dipole potential of the form~\cite{2011xix, Knobloch2017}
\begin{equation}
    U_{\rm el} (\rr) = - \frac{\alpha(0)}{2(4\pi\varepsilon_0)^2}\left( \sum_i \frac{q_i }{\left|\rr-\rr_i\right|^2} \right)^2\,,
    \label{eq:Uel}
\end{equation}
where $q_i$ is the effective charge of the $i$-th monolayer atom, and $\varepsilon_0$ is the vacuum permittivity.

\subsection{Repulsive potentials and absorption}
As the diffractive atom approaches the monolayer at interatomic distances, its electron orbitals begin to overlap with those of the monolayer atoms, triggering a strong repulsive potential due to Pauli repulsion. This repulsive interaction significantly impedes diffraction through the intact hBN monolayer, as well as through small holes {at thermal energies}~\cite{Brand2023, kanitz2024, Brand_2019}. The diffractive atoms experience a sharp change in their velocity.
We model this repulsion by adding a filter function to our propagation that removes any part of the wavefunction within the van der Waals radius of the monolayer atoms. The van der Waals radius of boron and nitrogen is {$1.92$\,{Å}}\cite{Mantina2009}, and {$1.55$\,Å}\cite{Bondi1964} respectively. 

\section{Diffraction modelling}
{Previous attempts at finding the diffraction patterns of the types of holes from Fig.\ref{fig:holes_shape} used the eikonal approximation and an atom-by-atom description of the hole reduction using the same potentials as this paper\cite{Osestad24}. Our new method improves upon this by collecting all the effects of the monolayer helium interactions into a single numerical simulation.} 
We find the diffraction patterns by {using the paraxial approximation,} simulating the transverse dynamics of quantum wavefronts as they pass through the holes. {Thus, we only simulate a {grid of a $15.9\times 15.9$Å $xy$}-plane and let the $z$-coordinate be a classical variable.} We subsequently apply Kirchhoff's diffraction formula to propagate the wavefront to an observation plane, located one metre away from the hBN plane. Fig.~\ref{fig:setup-illustration} illustrates how we move the transverse part of the wavefunction through the hole and onto the observation plane in Layer 4.

\subsection{Evolution of wave fronts}
The wavefunction plane, denoted by $\psi(\bm{r}, t)$, evolves according to the Schrödinger equation
\begin{equation}
    -\frac{\hbar^2}{2m}\nabla^2 \psi(\bm{r}, t)  + U(\bm{r},t) \psi(\bm{r},t) = \mi \hbar\frac{\partial\psi}{\partial t}(\bm{r}, t)\,.
\end{equation}
with $U(\bm{r}, t) = U_{\rm CP}(\bm{r},t) + U_{\rm el}(\bm{r},t)$.
We solve the propagation of the atom through the holes in the monolayer using the split operator method~\cite{Feit1982}. {We assume the wavefront is uniformly distributed over the simulated coordinates. Starting from $z_0 = -4.23$ {Å} away, we change the $z$-coordinate according to $z(t) = z_0 + vt$, which corresponds to the wavefront moving through the hole at a velocity $v$.} Once we reach a preset distance from the monolayer, where the potentials are once again negligible, we stop the simulation and use Eq.~\eqref{eq:kirchhoff} to find the diffraction pattern based on this final simulation plane. {More information on the simulation is found in the supplementary information. Different time steps were used for different velocities. The results were converged with respect to the time step.}

\subsection{Absorption of atoms} 
{Diffracting atoms that hit the monolayer atoms are scattered backwards, and therefore do not transmit through the holes. Relative to the propagating plane being modelled, any part of the matter-wave that is unable to transmit through the hole appears to be absorbed by the monolayer.} We model the absorption of the diffracted atoms by applying filter functions $F_i(r)$ to the atoms at each time step, defined by
\begin{equation}
F_i(r) = \begin{cases}
0 & \text{if } r \leq 0.8 \cdot r_{\rm vdW} \\
%\displaystyle 
\sin^4\left( \frac{\pi}{2}\frac{r - 0.8 \cdot r_{\rm vdW}}{0.2 \cdot r_{\rm vdW}}\right) & \text{if } 0.8 \cdot r_{\rm vdW} < r < r_{\rm vdW} \\
1 & \text{if } r \geq r_{\rm vdW}
\end{cases}
\label{eq:filter-function-cases}
\end{equation}
where $r_{\rm vdW}$ is the van der Waals radius of individual atoms in the hBN monolayer and $r$ is the separation distance between the helium atom and the $i$-th hBN atom. The full filter function is the product of all these individual filter functions
\begin{equation}
    F_{\rm full}(r) = \prod_i F_i(r)\,.
\end{equation}
The filter function removes the diffracting atoms that hit the monolayer, as these would not pass through the hole.
%\mw{What is happening then with the filter function? I.e. what is it doing?}
%\kz{where does the filter function appear in subsequent equations? In our previous paper, the transmission function/ filter function was then included in Eq. (26). Are we missing this term in Eq.~\ref{eq:kirchhoff}?}\eko{it doesent it is multiplied by $psi$ at each timestep, seperately. The effect is then encoded in $psi$ and we do not need }

\subsection{Evolution from monolayer to observation plane}
We assume that there is an observation plane located one metre from the monolayer. The wave front evolves freely to the observation plane after the atom passes through the hole, and is far enough away not to be affected by the monolayer. The resulting pattern at the observation plane can be found using the Kirchhoff diffraction formula\cite{BornWolfOptics,Brand15}
\begin{equation}
    \psi'(\bm{r}) = \frac{A k}{2\pi i}\int d^2 \rho\, \psi(\bm{\rho}) \frac{e^{\mi k s}}{2 s}\left[1 + \cos\theta\right]\,.
    \label{eq:kirchhoff}
\end{equation}
%\kz{Is this equation comparable to Eq. (26) is the other paper?}\eko{yes, they are very much the same, this one makes more assumptions}
Here, $A$ is the amplitude of the wave function, and $k = 2\pi / \lambda_{\rm dB}$. $s=|\bm{r} - \bm{\rho}|$ denotes the distance between the coordinates $\bm{\rho}$ of the final simulated plane (Layer 3 in Fig.~\ref{fig:setup-illustration}), and the coordinates $\bm{r}$ at which the diffraction pattern is measured (Layer 4 in Fig.~\ref{fig:setup-illustration}). $\theta$ is the angle between the normal to the monolayer and the vector $\bm{s}$. $\psi(\bm{\rho})$ is then the final state of the wave function as it passes the hole. {With our parameters, Eq.~\eqref{eq:kirchhoff} is a very good approximation of the Fourier transform of $\psi(\bm{\rho})$.} The probability of hitting the observation plane at $\bm{r}$ is then given by $|\psi'(\bm{r})|^2$. 

\section{Results}
We performed several numerical tests to calculate the transmission and diffraction of atoms through the hole geometries shown in Fig.~\ref{fig:holes_shape}. The new results differ from the semi-classical ones at all velocities, even though we would expect the results to converge for the highest velocities considered here. {
In the semi-classical model, the effective absorption radius was calculated using a hard-sphere model and the velocity of the incident atoms. 
In the new method, the absorption radii remain constant and the wavefront is shaped by the interactions between the diffracting atom and the atoms in the monolayer. This leads to significant differences in the shape of the hole and different results for the diffraction pattern at high velocities, even though the phase shifts converge.} At lower velocities, the newer and more complete quantum simulation leads to a largely increased transmission area. %At higher velocities we see more \kz{this sentence was unfinished}\eko{Dont remember what it was supposed to say} 

\subsection{Transmission through holes}
\begin{figure*}
    \centering
    \includegraphics[width = 0.8\textwidth]{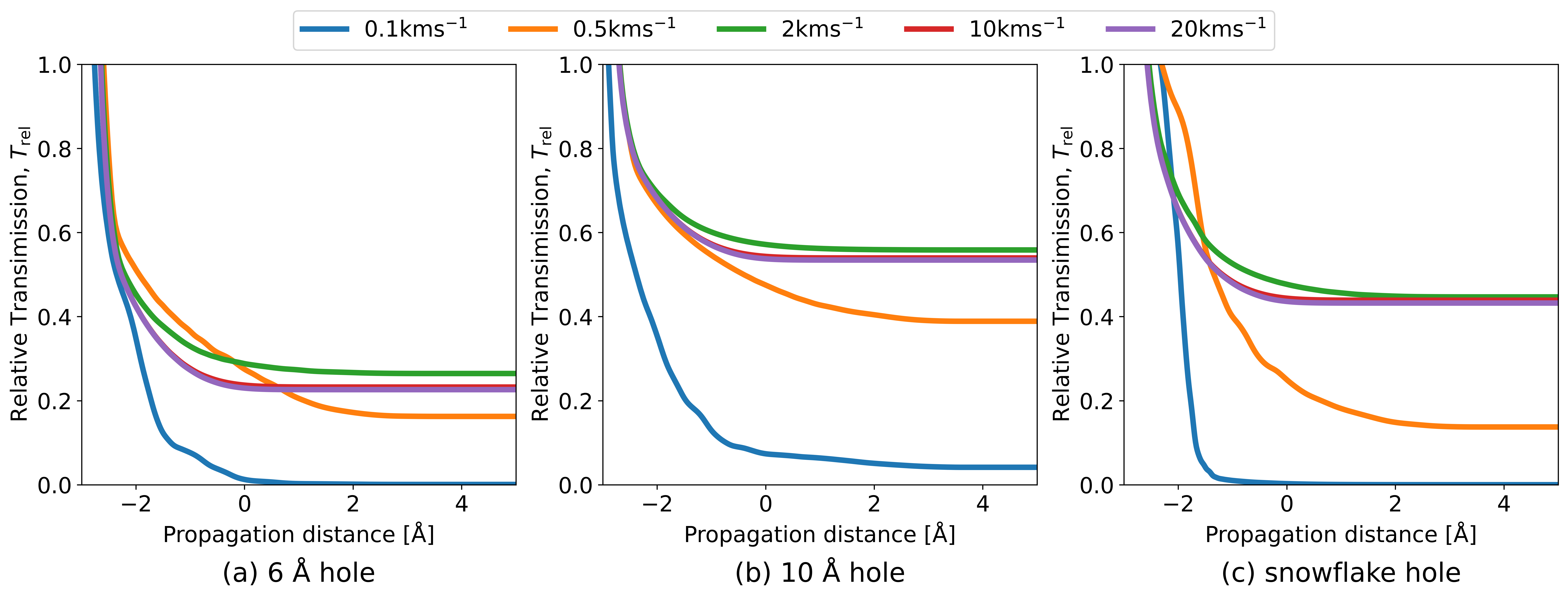}
    \caption{The relative transmission, $A_{\rm eff}$ [Eq.~\eqref{eq:effective-transmission-area}]{, normalized to the shaded areas in Fig.\ref{fig:holes_shape}}, for different velocities in dependence of the displacement from the hBN plane. The monolayer plane is {located} at {a propagation distance of} 0. {As the diffracted atoms pass the hole, more and more of the wavefunction enters the absorption area.}}
    \label{fig:transmission-distance}
\end{figure*}
\begin{figure}
    \centering
    \includegraphics[width=0.5\linewidth]{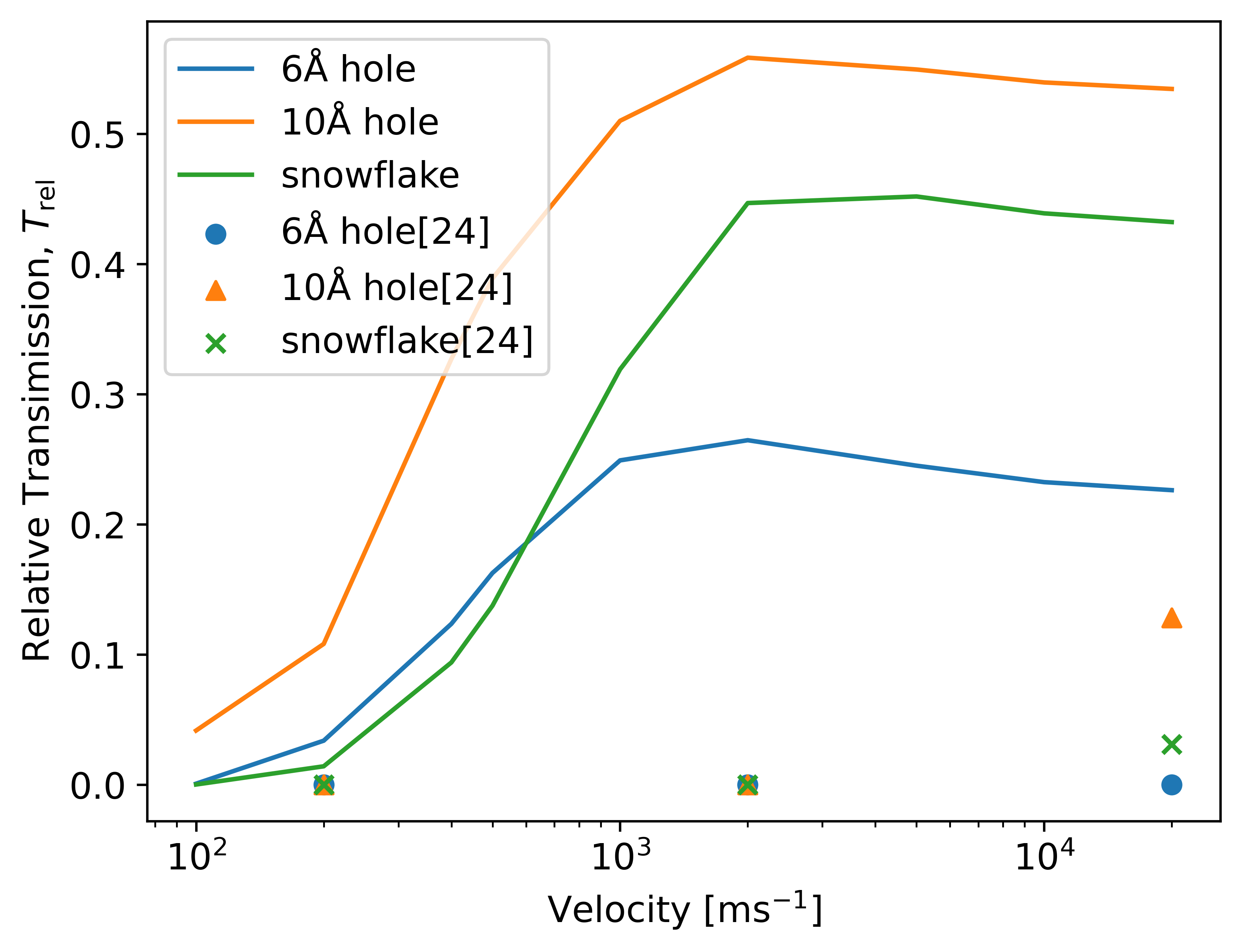}
    \caption{Final effective transmission area [Eq.~\eqref{eq:effective-transmission-area}]{, normalized to the shaded areas in Fig.\ref{fig:holes_shape},} in dependence  of the speed of the diffracted atom. Also included are the areas one gets based on the hole reduction calculated in the {\textit{s}emi-\textit{c}lassical [sc] approach\cite{Osestad24}, as shown with dots, triangles and crosses}.
    No transmission occurs at very slow velocities {below approx. 0.1 k\ms}. {The transmission} starts to increase with increasing velocity until it reaches a maximum and {flattens out}. {We do not observe this same effect for the semi-classical results, for the range of velocities shown here.}
    %\mw{I do not see the decrease, are you really sure about that?}
    The snowflake hole {(solid green line)} starts out with {a smaller transmission area} than the 6\,Å hole {(solid blue line)}, but eventually overtakes it. }
    \label{fig:transmission-speed}
\end{figure}

In the semi-classical approach, one models a reduction in the hole size due to the attractive forces between the membrane and the atom~\cite{Buhmann12a, Brand15, Fiedler2022, Osestad24}. This gives a velocity-dependent reduction in the effective size of the hole. In Fig.~\ref{fig:transmission-distance}, we see that we obtain the same type of velocity-dependent hole reduction with our method. Here, we define the {relative transmission, $T_{\rm rel}$,} as the relative amount of the wave function remaining compared {to how much could have transmitted through the central blue circles in Fig.~\ref{fig:holes_shape}, whitout van der Waals interactions.}

\begin{equation}
    \label{eq:effective-transmission-area}
    T_{\rm rel} = \frac{\int d^2r|\psi(\bm{r},t)|^2}{\int\limits_{A_{\rm circ}} d^2r |\psi(\bm{r},0)|^2}\, ,
\end{equation}
with {$A_{\rm circ}$ being the area of the central blue circles in Fig.~\ref{fig:holes_shape}. These circles are there to give us a real area to compare the transmission to}. In Fig.~\ref{fig:transmission-speed}, we see that for the very slowest atoms at around 0.1\,{k}\ms~almost no transmission occurs. Then the effective area increases with the velocity up to somewhere between 2\,{k}\ms~and 5\,{k}\ms. It then mostly flattens out{ as the velocity increases.}

\subsection{Diffraction patterns}
The diffraction pattern is found by taking the final state of the propagation simulation and evolving it according to Eq.~\eqref{eq:kirchhoff}. We evolve the wavefunction up to a distance of 1 metre from the hBN plane, since this distance is well in the far-field regime; the diffraction pattern is essentially the amplitude squared of the Fourier transform of the wave function after the hole~\cite{BornWolfOptics}. 

\begin{figure}
    \centering
    \includegraphics[width=0.75\linewidth]{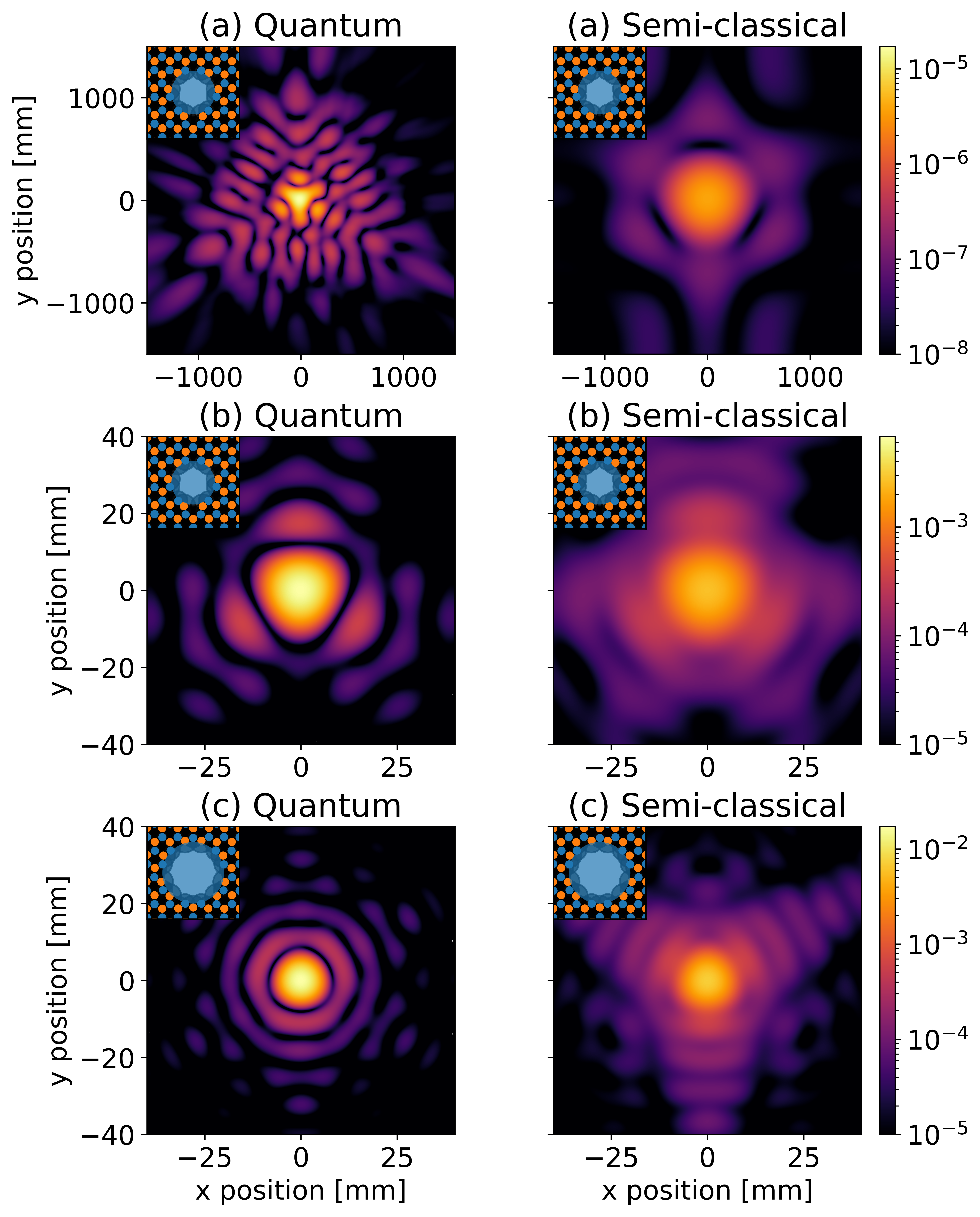}
    \caption{Comparison of the diffraction patterns, at a distance of 1 metre from the hole. The left column shows the results from our current quantum method, whereas the right column shows the results from the semi-classical approach~\cite{Osestad24}. Results for the snowflake hole [Fig. \ref{fig:holes_shape}(c)] at (a) 2\,{k}\ms, and (b) 20\,{k}\ms. (c) Results for the 10\,Å hole [Fig. \ref{fig:holes_shape}(b)] at 20\,{k}\ms. The colours represent the probabilities per mm$^{2}$ of finding the atom at $(x,y)$, plotted on a logarithmic scale.}
    \label{fig:diffraction}
\end{figure}

Fig. \ref{fig:diffraction} compares the diffraction patterns from our quantum method with our previous results obtained using a semi-classical method~\cite{Osestad24}. In part (a), we see this comparison for a helium matter-wave fired at the snowflake hole with a velocity of 2\,{k}\ms. {Here the atom is spread over a several m$^2$ area due to the transverse momentum imparted by the hole.} Fig. \ref{fig:diffraction}(a) shows that the angular resolution of the diffraction pattern is much greater in the quantum simulation compared to that in the semi-classical simulation. {This is due to the {strong van der Waals distortion of the wave front near the edges.}}

%There is a significant difference in how far the atoms are spread due to the large difference in the calculated effective transmission area. The larger transmission leads to less spread of the atoms compared to the older method. 
In Fig. \ref{fig:diffraction}(b) we see the same comparison, but for a helium matter with a velocity of 20\,{k}\ms. Here, the new method predicts a larger spread, despite the larger effective transmission area. The explanation of this is that the Casimir--Polder force acts more strongly closer to the monolayer atoms. 
{The new method predicts a shape for the holes that allows the diffracting atoms to pass closer to the monolayer, thus} they can obtain more transverse velocity when passing the hole {as the transverse forces are stronger near the edges of the hole.}
The same effect is found in Fig. \ref{fig:diffraction}(c), where the 20\,{k}\ms~case is plotted for the 10\,Å hole. In addition, this particular example shows how much smoother the diffraction pattern becomes with the current method being very close to rotationally symmetric.

\section{Conclusion}
In this paper, we applied a new method of modelling the diffraction of atoms by sub-nanometre holes in hBN, challenging the conventional semi-classical approach and demonstrating the need for a quantum-mechanical treatment. By applying a split-operator numerical approach to the transverse component of an atom's quantum mechanical wave function, we have accurately modelled the diffraction process.
%We have demonstrated the limits of the semi-classical approach that has been used previously, and therefore shown that a more full quantum-mechanical approach is necessary for describing diffraction of atoms at these sub-nanometre holes. 
%We have applied a split-operator numerical approach to the transverse component of an atoms quantum mechanical wave function as it moves through holes in hexagonal boron nitride. 
For each time step, we updated the potential, which corresponds to the propagation of the wavefront through the hBN monolayer. Once the propagation through the hole is complete, we find the far-field diffraction pattern of the hole, and compare our results to the previous attempt to calculate this in Ref. \cite{Osestad24}.  

We predict that diffraction is possible for much smaller holes than previously thought, and that the effective area of the hole through which atoms can diffract is much larger, particularly for lower velocities. This discovery has significant implications for the development of new technologies that rely on atomic-scale manipulation, such as atom interferometry sensing and atomic holography.

Furthermore, our findings highlight the importance of considering the quantum-mechanical nature of atomic diffraction, even at high velocities. This is particularly evident in the effects related to dispersion forces, where the velocity and trajectory of the helium atoms depend on how close they pass to the edge of the hole. This can lead to a large velocity spread, leading to reduced angular resolution in the diffraction patterns. 
%There are also effects related to the dispersion forces creating a larger velocity spread for higher velocity diffraction due to the diffracting atoms ability to get closer to the monolayer.
%These effects show that a quantum-mechanical approach is necessary to describe the diffraction of atoms 
%through these small holes accurately even at high velocities. 

For future work, it would be interesting to study the effects of the wave packet velocity changing in the propagation direction near the mask, especially for slower atoms. Investigations into other materials, such as graphene, or using other atoms or molecules, could also be beneficial. Testing what happens with large incidence angles would also be interesting. {Our method also potentially {becomes inaccurate} when it comes to inelastic scattering due to phonon and plasmon excitations of the monolayer{, which has been studied before in graphene\cite{Brand2023}}.  In future work, we will address these effects as potential energy and coherence losses.}

%Bibliography
\bibliographystyle{unsrt}  
\bibliography{references}

\end{document}